\documentclass[amssymb,aps,prl,twocolumn,groupedaddress]{revtex4-1}

\usepackage[]{amsmath}
\usepackage[]{bbm}
\usepackage[]{color}
\usepackage[version=3]{mhchem}
\usepackage[utf8x]{inputenc}
\usepackage[]{graphicx}
\usepackage[]{hyperref}

\newcommand{\ctskappa}{$\kappa$-(BEDT-TTF)$_2$Hg(SCN)$_2$Cl}
\newcommand{\ctstmttf}{(TMTTF)$_2$SbF$_6$}
\begin{document}

\title{Determination of magnetic form factors for
organic charge transfer salts: a first principles investigation}

\author{Francesc Salvat-Pujol}
\affiliation{Institut f\"ur Theoretische Physik, Goethe-Universit\"at
Frankfurt, Max-Von-Laue-Stra{\ss}e 1, 60438 Frankfurt am Main, Germany}

\author{Harald O. Jeschke}
\email[Email:]{jeschke@itp.uni-frankfurt.de}
\affiliation{Institut f\"ur Theoretische Physik, Goethe-Universit\"at
Frankfurt, Max-Von-Laue-Stra{\ss}e 1, 60438 Frankfurt am Main, Germany}

\author{Roser Valent{\'\i}}
\affiliation{Institut f\"ur Theoretische Physik, Goethe-Universit\"at
Frankfurt, Max-Von-Laue-Stra{\ss}e 1, 60438 Frankfurt am Main, Germany}

\begin{abstract}
  Organic charge transfer salts show a variety of complex phases
  ranging from antiferromagnetic long-range order, spin liquid, bad
  metal or even superconductivity.  A powerful method to investigate
  magnetism is spin-polarized inelastic neutron scattering. However,
  such measurements have often been hindered in the past by the small
  size of available crystals as well as by the fact that the spin in
  these materials is distributed over molecular rather than atomic
  orbitals and good estimates for the magnetic form factors are
  missing.  By considering Wannier functions obtained from density
  functional theory calculations, we derive magnetic form factors for
  a number of representative organic molecules. Compared to Cu$^{2+}$,
  the form factors $|F(\mathbf{q})|^2$ fall off more rapidly as
  function of $q$ reflecting the fact that the spin density is very
  extended in real space. Form factors
  $|F(\mathbf{q})|^2$ for TMTTF, BEDT-TTF and (BEDT-TTF)$_2$ have
  anisotropic and nonmonotonic structure.
\end{abstract}

\maketitle
\newpage

Since the discovery of superconductivity in the Bechgaard salt
(TMTSF)$_2$PF$_6$~\cite{Jerome1980}, the complex phase diagrams of
organic charge transfer salts have inspired intense research
efforts~\cite{Toyota2007}. Among the families of charge transfer salts
with magnetic and superconducting phases, the more one-dimensional
Fabre salts~\cite{Ardavan2012} and the more two-dimensional salts
based on BEDT-TTF molecules in $\kappa$-type structural
arrangement~\cite{Wosnitza2007,Kanoda2011} have attracted a lot of
attention.  Within the many experimental techniques used to study
these organic materials, magnetic inelastic neutron scattering has, to
our knowledge, so far not been used. This technique has played an
outstanding role in the investigation of cuprate high temperature
superconductors~\cite{Fong1999,Tranquada2004} and its application to
organics would mean a significant progress~\cite{Ardavan2012}. The
size of available crystals have limited the application of neutron
techniques on charge transfer salts, and only the phonon response of a
few materials like
$\kappa$-(BEDT-TTF)$_2$Cu(SCN)$_2$~\cite{Pintschovius1997} has been
studied by inelastic neutron scattering (INS). For the quantitative
interpretation of magnetic inelastic neutron scattering spectra,
however, besides significant crystal sizes the knowledge of the
magnetic form factor is necessary. In magnetically ordered organic
charge transfer salts, the polarized neutrons are scattered by spins
which are not localized on atomic-like Cu $3d_{x^2-y^2}$ orbitals as
in cuprates but spin densities which are distributed over extended
molecular orbitals.  While atomic magnetic form factors are
tabulated~\cite{ITC}, magnetic form factors for molecular orbitals are
often not known. Due to the large spatial extension and inhomogeneity
of a molecular orbital, the corresponding magnetic form factor can be
expected to exhibit more structure than their atomic
counterparts. Walters {\it et al.}~\cite{Walters2009} have
demonstrated for the one-dimensional cuprate Sr$_2$CuO$_3$ that
structure in the spin density distribution beyond the regular Cu
$3d_{x^2-y^2}$ shape has important consequences for the quantitative
evaluation of magnetic INS.  In this work, we will extend this
approach to the molecular orbitals carrying the spin in one- and
two-dimensional charge transfer salts. We investigate two
representative examples, {\ctstmttf} (where TMTTF stands for
tetramethyl-tetrathiafulvalene) and {\ctskappa} (where BEDT-TTF
denotes bis-(ethylenedithio)-tetrathiafulvalene).




{\it Method.-} In magnetic neutron scattering, neutrons are used to
probe the spin density of a material. For a given momentum transfer
$\mathbf{q}=\mathbf{k}-\mathbf{k'}$ and energy transfer
$\hbar\omega=\frac{\hbar^2}{2m}(k^2-{k'}^2)$, the magnetic scattering
cross section is given as~\cite{Furrer2009}
\begin{equation}\begin{split}
    \frac{d^2\sigma}{d\Omega d\omega} &= (\gamma r_0)^2 \frac{k'}{k}
    |F(\mathbf{q})|^2 e^{-2W(\mathbf{q})}\times
    \\&\times\quad\sum_{\alpha\beta} \bigg(\delta_{\alpha\beta} -
    \frac{q_\alpha q_\beta }{q^2}\bigg)
    S^{\alpha\beta}(\mathbf{q},\omega)
\end{split}\end{equation}
with Debye-Waller factor $e^{-2W(\mathbf{q})}$ and the magnetic
scattering function $S^{\alpha\beta}(\mathbf{q},\omega)$. Here, we
focus on the static magnetic form factor $F(\mathbf{q})$. This
quantity is defined as the Fourier transform of the electronic spin
density $\rho_s(\mathbf{r})$,
\begin{equation}
  F(\mathbf{q})  =
  \int{\rm d}^3r\;
  e^{{\rm i}\mathbf{q}\cdot\mathbf{r}}
  \rho_s(\mathbf{r})\,.
  \label{}
\end{equation}
The electronic spin density is given by
\begin{equation}
  \rho_s(\mathbf{r})=\rho_\uparrow(\mathbf{r})-\rho_\downarrow(\mathbf{r})\,,
  \label{}
\end{equation}
where $\rho_\uparrow(\mathbf{r})$ and $\rho_\downarrow(\mathbf{r})$
denote the electronic density with spin up and down, respectively, of
the local scatterer. For crystalline systems, these spin densities can
be obtained from first-principles solid-state calculations with
density-functional theory (DFT). However, within DFT, the eigenstates
of the system are Bloch states (characterized by a band index, a
wavevector, and spin) which are periodically extended waves. In order
to obtain a spin density localized on a given scatterer (atom or
molecule), a projection of the Bloch state onto a state localized at
the corresponding scatterer must be carried out. The resulting
localized orbitals are so-called Wannier orbitals. Once the localized
orbitals are known, obtaining the spin density and the magnetic form
factor result from evaluating numerically the expressions given above.

\begin{figure}
  \centering
  \includegraphics[width=0.47\textwidth]{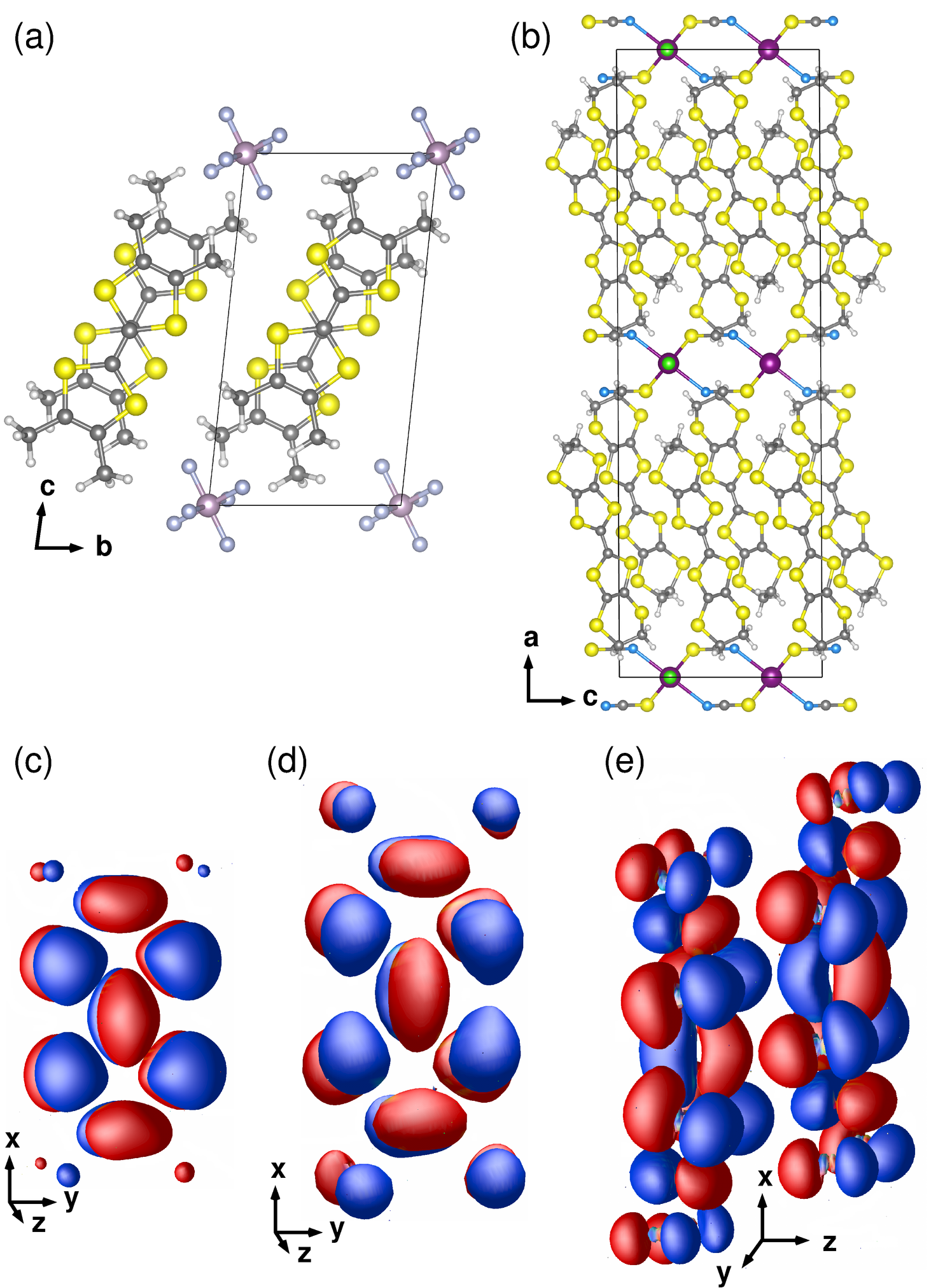}
  \caption{ (Color online) Crystal structures for (a) {\ctstmttf} and
    (b) {\ctskappa}.  Wannier orbitals for (c) the TMTTF molecule, (d)
    the BEDT-TTF molecule and (e) the (BEDT-TTF)$_2$ dimer, calculated
    within the crystal structures in (a) and (b), respectively.}
  \label{fig:structwann}
\end{figure}

\begin{figure}
  \centering
  \includegraphics[width=0.47\textwidth]{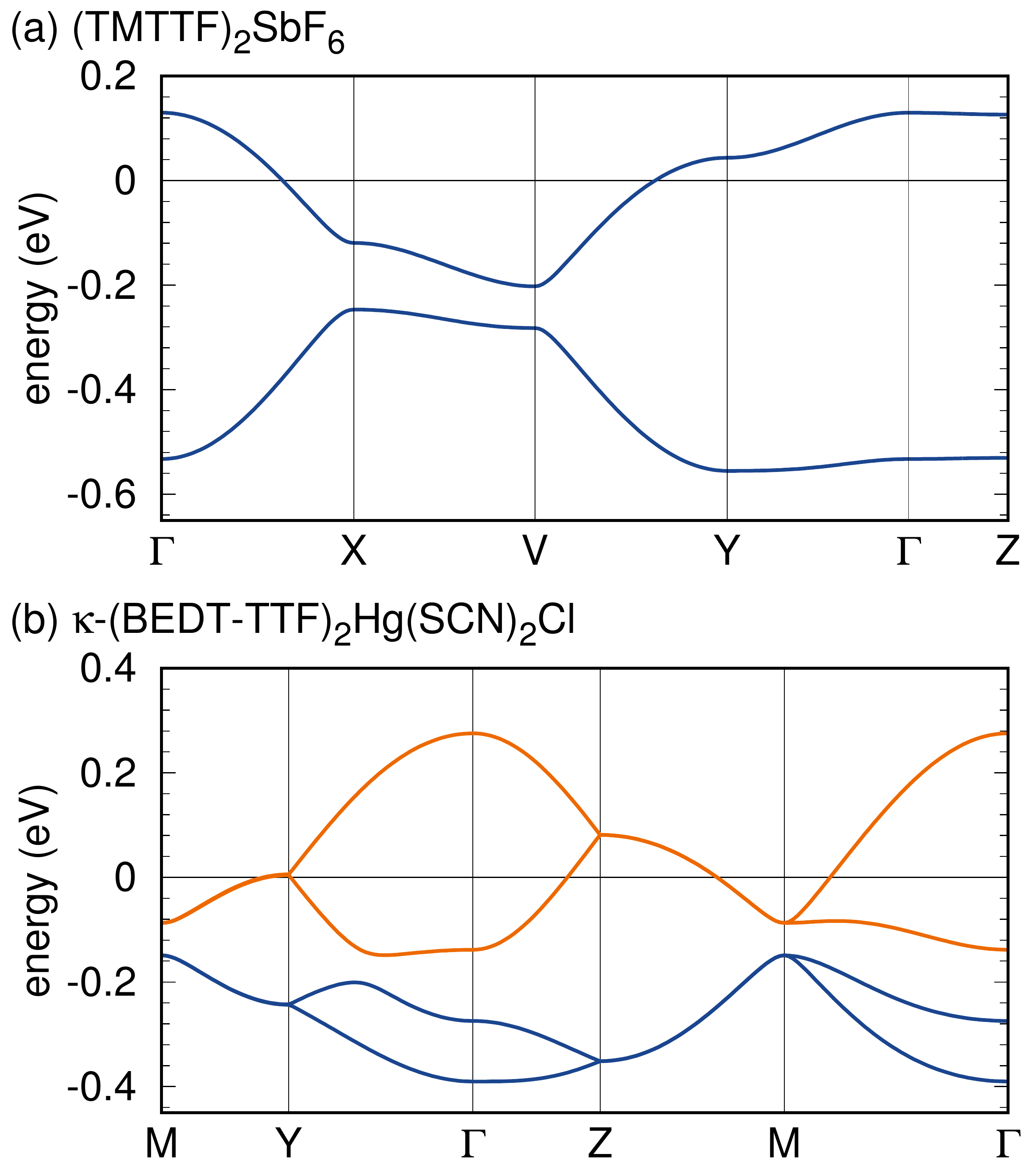}
  \caption{ (Color online) Bandstructures near $E_{\rm F}$ for (a)
    {\ctstmttf} and (b) {\ctskappa}. The two bands in (a) originate
    from the highest occupied molecular orbitals of the two TMTTF
    molecules in the unit cell, the four bands in (b) from the highest
    occupied molecular orbitals of the four BEDT-TTF molecules. }
  \label{fig:bandstructure}
\end{figure}

\begin{figure}
  \centering
  \includegraphics[width=0.47\textwidth]{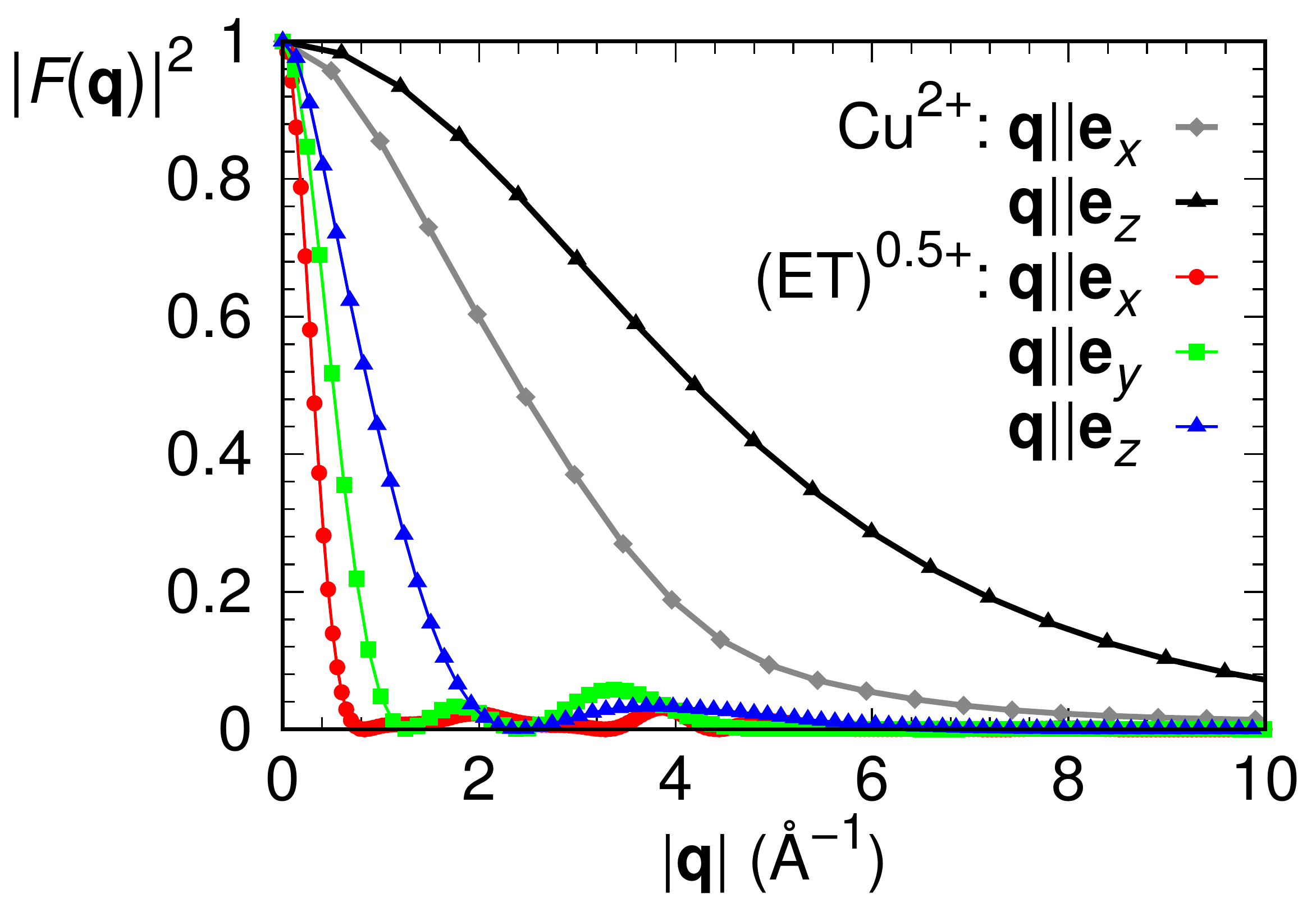}
  \caption{ (Color online) Comparison between magnetic form factors
    for the Cu$^{2+}$ ion and the (BEDT-TTF)$^{0.5+}$ ion (BEDT-TTF is
    abbreviated further as ET). }
  \label{fig:fqcu}
\end{figure}

\begin{figure*}
  \centering
  \includegraphics[width=0.8\textwidth]{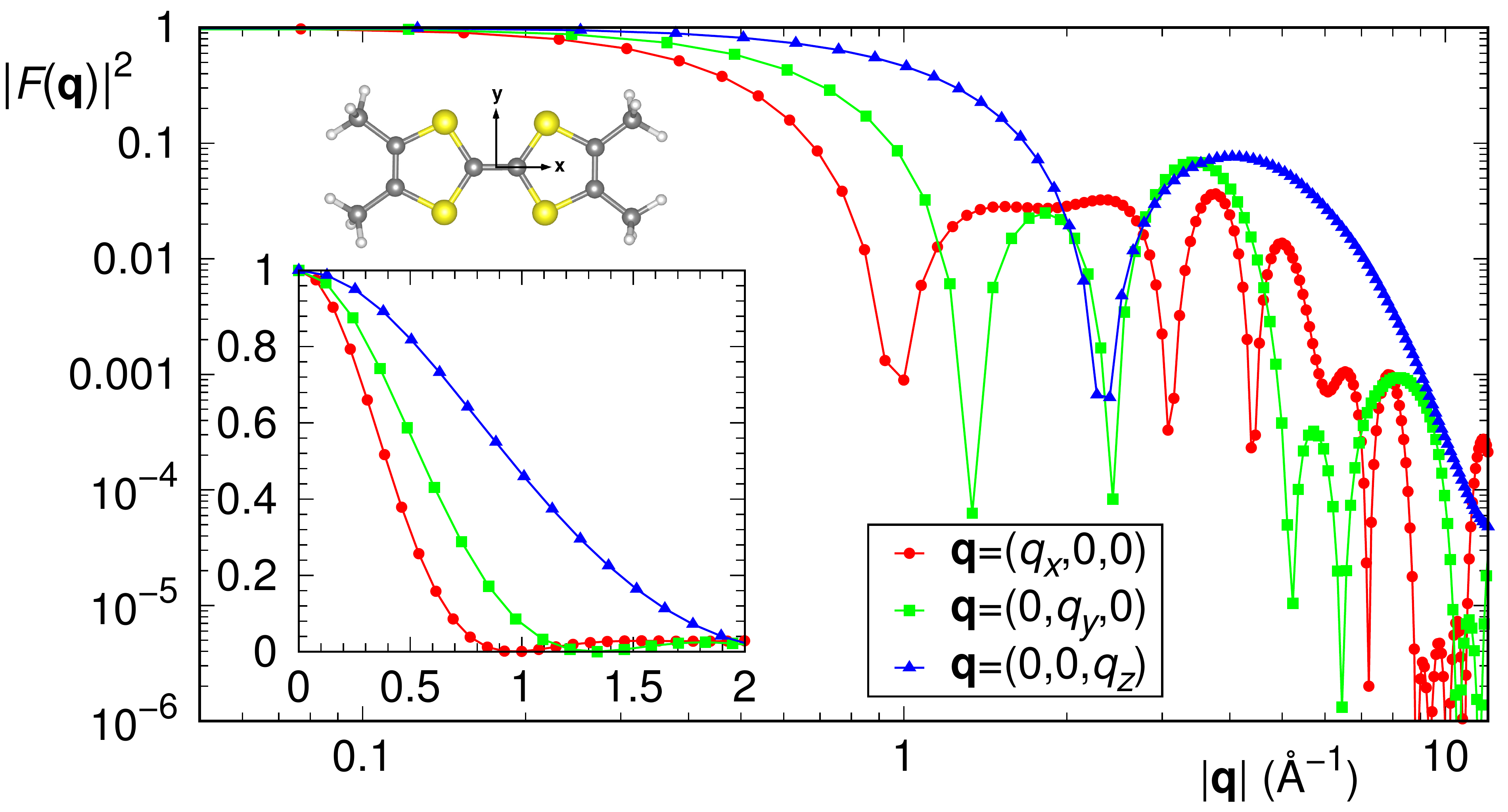}
  \caption{Magnetic form factor for {\ctstmttf}; the inset shows the
    same data in linear scale. }
  \label{fig:fqtmttf}
\end{figure*}

\begin{figure*}
  \centering
  \includegraphics[width=0.8\textwidth]{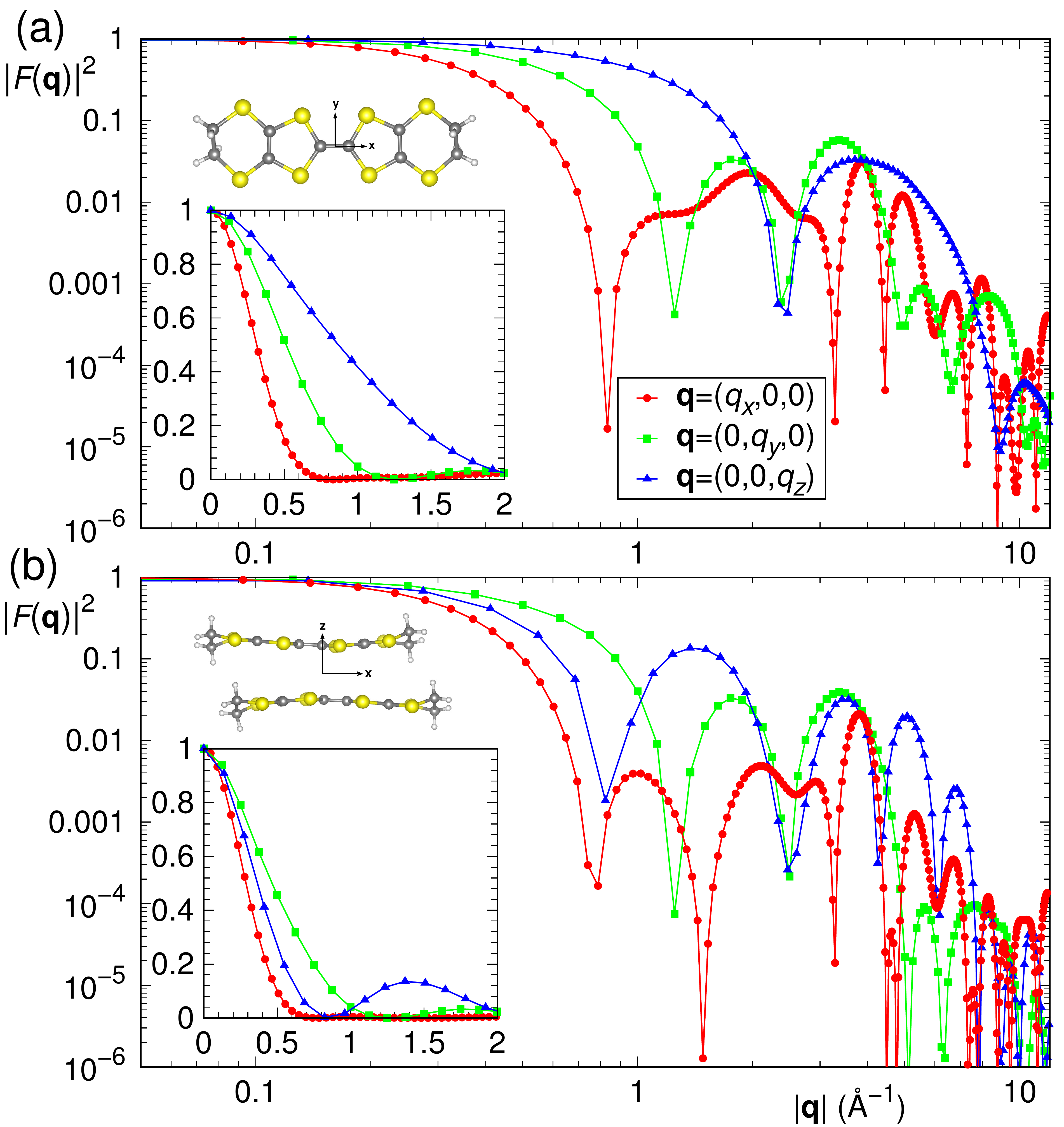}
  \caption{Magnetic form factors for {\ctskappa}. (a) Form factor for
    a BEDT-TTF molecule, and (b) form factor for a (BEDT-TTF)$_2$
    dimer.}
  \label{fig:fqkappa}
\end{figure*}

Density functional theory calculations are performed on the
full-potential non-orthogonal local-orbital basis set, as implemented
in the FPLO code~\cite{Koepernik1999} and the generalized gradient
approximation~\cite{perdew1996} to the exchange-correlation functional
is adopted.  It should be mentioned that, although one should in
principle carry out spin-polarized DFT calculations for half-filled
systems like those examined in this work, it is permissible to carry
out a non-magnetic calculation and consider the unpaired electron
occupying the half-filled band at the Fermi level as giving rise to
the net spin density~\cite{Walters2009}. To perform the projection of
the (Kohn-Sham) Bloch states on a localized orbital, we use the
projective Wannier functions within the FPLO basis as described in
Ref.~\cite{Eschrig2009}. The magnetic form factor is computed from the
resulting spin density by means of the fast Fourier transform (FFT).

{\it Results.-} Magnetic form factors have been calculated for two
representative organic charge transfer salts, {\ctskappa} and
{\ctstmttf}; the crystal structure of these systems are taken from
Refs.~\cite{Jacko2013} and \cite{Drichko2014}, respectively, and are
displayed in panels (a) and (b) of Fig.~\ref{fig:structwann}.  Whereas
the C and S atoms in {\ctstmttf} are coplanar, {\ctskappa} exhibits a
minor non-coplanarity along the main axis of the molecule. This
non-coplanarity is accentuated by the two ethylene end groups, which
can have socalled eclipsed and staggered out of plane
twists~\cite{Toyota2007}.  DFT calculations were performed on $8\times
8\times 8$ and $6\times 6\times 6$ $k$ meshes for {\ctstmttf} and
{\ctskappa}, respectively.
For {\ctstmttf}, the two bands near the Fermi level shown in
Fig.~\ref{fig:bandstructure}~(a) (compare also Ref.~\cite{Jacko2013})
are represented by two molecular Wannier functions, one of which is
shown in Fig.~\ref{fig:structwann}~(c). For {\ctskappa}, the four
bands formed by the highest occupied molecular orbital states of the
four BEDT-TTF molecules in the unit cell (see
Fig.~\ref{fig:bandstructure}~(b)), are represented by four Wannier
functions like the one shown in Fig.~\ref{fig:structwann}~(d).  A
system of Cartesian coordinates $x$, $y$, and $z$ is introduced on the
molecule in such a way that $x$ points along the long axis of the
molecule, $y$ is on the molecule pointing along the shorter axis and
$z$ is perpendicular to the molecule.

First, we compare in Fig.~\ref{fig:fqcu} the magnetic form factor for
the Cu$^{2+}$ ion in Sr$_2$CuO$_3$~\cite{Walters2009} with the
magnetic form factor for the (BEDT-TTF)$^{0.5+}$ ion in
{\ctskappa}. Due to the larger spatial extent of the BEDT-TTF Wannier
function (see Fig.~\ref{fig:structwann}~(d)) compared to the Cu
$3d_{x^2-y^2}$ Wannier function, the magnetic form factor of
(BEDT-TTF)$^{0.5+}$ drops to its first minimum at much lower $q$
values.  Figures \ref{fig:fqtmttf} and \ref{fig:fqkappa}~(a) display
the magnetic static form factors for TMTTF molecules in {\ctstmttf}
and for BEDT-TTF molecules in {\ctskappa}, respectively, as a function
of $q_x$, $q_y$, and $q_z$, the Fourier-conjugate variables of $x$,
$y$, and $z$. As can be seen in Fig.~\ref{fig:structwann}, the charge
densities are broad along $x$ and $y$ and rather concentrated along
$z$. Accordingly, the form factors are comparatively narrow for $q_x$
and $q_y$ and broader for $q_z$. Note that, as opposed to the case of
a free atom, the form factor is not a steadily decreasing function: it
exhibits marked features. These features reflect the fact that the
charge density is strongly modulated over the region in space occupied
by the molecule.

Due to the fact that both {\ctstmttf} and {\ctskappa} are half-filled
systems if we focus on the antibonding bands arising from the highest
occupied molecular orbitals only, the dimers (TMTTF)$_2$ and
(BEDT-TTF)$_2$, each hosting one hole, can be considered spin 1/2
objects.  In order to aid in the interpretation of magnetic inelastic
neutron scattering data, we also provide in Fig.~\ref{fig:fqkappa}~(b)
the magnetic form factor associated to a BEDT-TTF dimer. This
corresponds to the dimer Wannier function shown in
Fig.~\ref{fig:structwann}~(e). The comparison of this form factor with
the form factor for an BEDT-TTF molecule shows that the main peak
along $q_z$ becomes narrower (since the spin density is broader along
$z$ for the dimer). Also the peak along $q_x$ becomes slightly
narrower and its fine-grained structure is also affected, owing to the
fact that the dimers are slightly shifted along $x$.

{\it Conclusions.-} By considering a combination of density functional
theory calculations, Wannier function construction and numerical
Fourier transformations, we have been able to derive accurate form
factors for organic molecules in crystalline systems. Such form
factors are indispensable for a quantitative analysis of magnetic
inelastic neutron scattering experiments of organic materials.  We
observe a number of differences between form factors for organic
molecules and transition metal ions: (i) due to the large spatial
extent of the spin density, the form factor for organic molecules
falls to its first minimum at much smaller $q$ values, (ii) due to
spatial modulation of the spin density, the form factors show richer
$q$-dependent structure, and (iii) the real space shape of the
molecular spin density leads to very anisotropic form factors.  These
consequences of the extended inhomogeneous molecular spin densities
can only be captured by accurate first principles calculations.  We
hope that this work will help further investigations of the behavior
of organic crystals with inelastic neutron scattering experiments.

\begin{acknowledgments}

  We would like to thank M. Medarde, S. T{\'o}th and Ch. R\"uegg for
  useful discussions and C. Broholm for pointing the problem out to
  us.  F.S.-P. gratefully acknowledges the support of the Alexander
  von Humboldt Foundation through a Humboldt Research Fellowship.
  H.O.J. and R.V.  thank the Deutsche Forschungsgemeinschaft for
  financial support through grant SFB/TR49.

\end{acknowledgments}

\end{document}